\documentclass[twocolumn,aip,reprint,amsmath,amssymb]{revtex4-2}
\usepackage[hidelinks]{hyperref}
\usepackage{graphicx}
\usepackage{epstopdf}
\usepackage{color}
\usepackage{bm}
\usepackage{amssymb}
\usepackage{amsfonts}
\usepackage{amsmath}
\usepackage{float}
\usepackage{siunitx}
\usepackage{multirow}

\begin{document}

\title{Strain effect on optical properties and quantum weight of monolayer MnBi$_2$X$_4$ (X = Te, Se, S)}

\author{Nguyen Tuan Hung}
\thanks{Author to whom correspondence should be addressed: nguyen.tuan.hung.e4@tohoku.ac.jp}
\affiliation{Department of Materials Science and Engineering, National Taiwan University, Taipei 10617, Taiwan}
\affiliation{Frontier Research Institute for Interdisciplinary Sciences, Tohoku University, Sendai 980-8578, Japan}
\author{Vuong Van Thanh}
\email{thanh.vuongvan@hust.edu.vn}
\affiliation{School of Mechanical Engineering, Hanoi University of Science and Technology, Hanoi 100000, Viet Nam}
\author{Mingda Li}
\affiliation{Quantum Measurement Group, MIT, Cambridge, MA 02139-4307, USA}
\affiliation{Department of Nuclear Science and Engineering, MIT, Cambridge, MA 02139-4307, USA}
\author{Takahiro Shimada}
\affiliation{Department of Mechanical Engineering and Science, Kyoto University, Kyoto 615-8540, Japan}

\date{\today}

\begin{abstract}
Manipulating the optical and quantum properties of two-dimensional (2D) materials through strain engineering is not only fundamentally interesting but also provides significant benefits across various applications. In this work, we employ first-principles calculations to investigate the effects of strain on the magnetic and optical properties of the monolayer MnBi$_2$X$_4$ (X = Te, Se, S). Our results indicate that biaxial strain enhances the Mn magnetic moment, while uniaxial strains reduce it. Significantly, the strain-dependent behavior, quantified through the quantum weight, can be leveraged to control the system’s quantum geometry and topological features. Particularly, uniaxial strains reduce the quantum weight and introduce anisotropy, thus providing an additional degree of freedom to tailor device functionalities. Finally, by analyzing chemical bonds under various strain directions, we elucidate how the intrinsic ductile or brittle fracture behavior of MnBi$_2$X$_4$ could impact fabrication protocols and structural stability. These insights pave the way for strain-based approaches to optimize the quantum properties in 2D magnetic topological insulators in practical device contexts.
\end{abstract}

\maketitle

The emergence of two-dimensional (2D) magnetism in topological materials creates a platform for spin-based applications in spintronics~\cite{vzutic2004spintronics}, multiferroics~\cite{cheong2007multiferroics}, and magnetic memory. The 2D topological materials are known for their characteristic of topologically protected quantum states~\cite{hasan2010colloquium,qi2011topological}. These quantum states are robust against local perturbations such as changes in size and temperature, making electronic devices based on 2D topological materials extremely stable in the environment. Moreover, the magnetic order can break the time-reversal symmetry of the 2D topological material~\cite{hung2024symmetry} and enhance the quantum nature of the 2D topological material~\cite{qi2008topological}. For example, Lee \textit{et. al.}~\cite{lee2021magnetic} observed a large second harmonic generation effect of the CrSBr monolayer below 146 K due to the phase transition from paramagnetic to ferromagnetic ordering. The energy band-gap opening in magnetism topological insulators with chiral edge states also gives rise to a quantum anomalous Hall (QAH) effect when the Fermi energy is localized inside the band gap~\cite{chang2013experimental,yu2010quantized}. The unique properties of the 2D magnetism topological insulators thus lead to advancements in topological device applications~\cite{tokura2019magnetic}.

The 2D material MnBi$_2$Te$_4$ is one of the unique layered structures that exhibit both non-trivial topology and intrinsic magnetism~\cite{li2019intrinsic,trang2021crossover,deng2020quantum,ren2022quantum,li2021electronic}. Similar 2D materials, such as MnBi$_2$Se$_4$ and MnBi$_2$S$_4$, have also been synthesized and theoretically predicted as promising candidates for exploring topological quantum states~\cite{zhu2021synthesis,nowka2017chemical,zhang2021tunable}. To investigate the quantum characteristics of these materials, researchers often employ Hall effect and magnetoresistance measurements to detect the QAH effect~\cite{chang2013experimental,yu2010quantized}, or they calculate the Berry curvature to determine the Chern number~\cite{zhu2022high,bosnar2023high}. However, these methods can be sensitive to sample quality and may require significant computational resources to yield accurate results. An alternative approach is to utilize quantum weight as a means to assess the quantum level of the system~\cite{onishi2024fundamental}. Quantum weight, which can be derived from the real part of optical conductivity using the $f$-sum rule~\cite{hung2024universal}, is directly related to the ground-state quantum geometry of the material~\cite{das2023intrinsic,kaplan2024unification,wang2023quantum}. Thus, the quantum weight can be a crucial parameter for quantifying the degree of quantumness of an insulating material~\cite{ghosh2024probing}. The direct correlation between quantum weight and optical properties provides a unique opportunity to manipulate the quantum characteristics of materials by adjusting their optical features. Strain engineering has emerged as an effective approach to tuning the optical properties of 2D materials~\cite{van2020first,van2022effects,hung2024symmetry}. Experimentally, strain can be applied biaxially or uniaxially using various techniques such as substrate bending or stretching, external pressure, or atomic force microscopy (AFM) tips~\cite{dai2019strain,wu2022strain,zhang20202d}. Therefore, understanding the strain effect on the optical properties and quantum weight of 2D topological insulators could open new ways for advancements in strain engineering to effectively control quantum properties.

\begin{figure*}[t]
    \centering
    \includegraphics[width=0.85\linewidth]{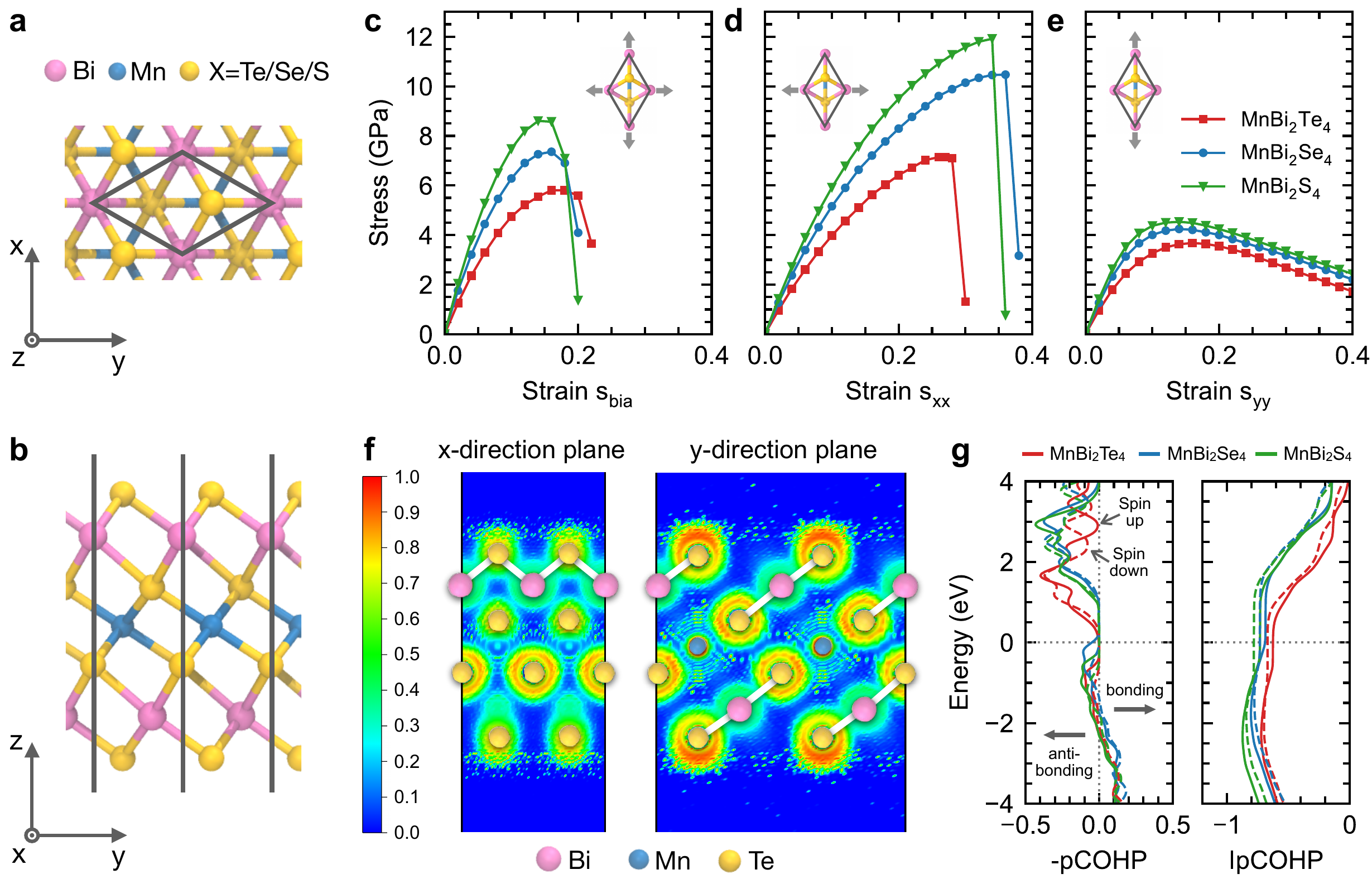}
    \caption{(a) Top view and (b) side view of the MnBi$_2$X$_4$ (X = Te/Se/S) structure, in which the solid box indicates the unit-cell. (c)-(e) The strain-stress curves of MnBi$_2$X$_4$ for biaxial strain ($s_{\text{bia}}$) and uniaxial strains along the $x$- ($s_{xx}$) and $y$-directions ($s_{yy}$), respectively. (f) Electron localization function in the $x$- and $y$-direction planes, in which the isosurface value is set at 1.0. (g) Projected crystal orbital Hamilton population (pCOHP) for spin-up (solid lines) and spin-down (dashed lines).}
    \label{fig:strain-stress}
\end{figure*}

In this letter, we explored the strain effect on the electronic, magnetic, and optical properties of the monolayer MnBi$_2$X$_4$ (X = Te, Se, S) using first-principles calculations. We found that the strain reduces and induces anisotropic behavior in the quantum weight.

The mechanical, electronic, and magnetic properties are calculated using the Quantum ESPRESSO package~\cite{giannozzi2009quantum}. We use a cutoff energy for plane-wave of 80 Ry and $\bm{k}$-points of $12 \times 12 \times 1$ and $24 \times 24 \times 1$ are used for the self-consistent field (SCF) and non-SCF calculations, respectively. The pseudopotentials of atoms are selected from the PS library~\cite{dal2014pseudopotentials}. A Hubbard $U$ correction is applied to the Mn-$3d$ orbitals to account for strong correlation effects. A Hubbard $U$-value of 4 eV is used for all calculations, chosen by comparing our computed energy gap with the reported experimental data~\cite{trang2021crossover}, as shown in Fig. S1 in Supporting Information. The dielectric functions and optical properties are calculated using the Yambo code~\cite{marini2009yambo} within the independent particle approximation. The slab models of MnBi$_2$X$_4$ are built with a 40 \AA\ vacuum along the $z$-direction to avoid the interactions between the slabs. All lattice constants and the positions of atoms (without and with strain) are optimized by the Broyden–Fletcher–Goldfarb–Shanno (BFGS) quasi-Newton algorithm~\cite{hung2022quantum}, in which the convergence values for the forces and stress components are 0.0001 Ry/a.u.$^3$ and 0.005 GPa, respectively. We apply an in-plane strain along two perpendicular directions $s_\text{bia}$, along only $x$-direction, $s_{xx}$, and along only $y$-direction, $s_{yy}$, with a step strain of 0.02 (or 2\%). The stress conditions ($\sigma_{xx}=\sigma_{yy}\neq 0$), ($\sigma_{xx}\neq 0, \sigma_{yy}=0$), and ($\sigma_{xx} = 0, \sigma_{yy}\neq 0$) are applied for $s_\text{bia}$, $s_{xx}$, and $s_{yy}$, respectively.

\begin{table*}[t]
\caption{In-plane lattice constant $a$ (\AA), effective thickness $h$ (\AA),  elastic constants $C_{ii}$ (GPa), Young modulus $Y$ (GPa), Poisson ratio $r$, critical strain $s^c$, and ideal strength $\sigma^i$ (GPa) of MnBi$_2$X$_4$ (X = Te/Se/S).}
\centering    
{\renewcommand{\arraystretch}{1.5} 
\setlength{\tabcolsep}{6pt}
\small
\begin{tabular}{cccccccccccccc}
\hline\hline
Structure & $a$ & $h$ & $C_{11}$ & $C_{12}$ & $C_{66}$ & $Y$ & $r$ & $s^c_{\text{bia}}$ & $s^c_{xx}$ & $s^c_{yy}$ & $\sigma^i_{\text{bia}}$ & $\sigma^i_{xx}$ & $\sigma^i_{yy}$\\ 
\hline    
MnBi$_2$Te$_4$ & 4.36 & 13.70 & 56.17 & 13.54 & 21.32 & 52.91 & 0.24 & 0.16 & 0.27 & 0.16 & 5.81 & 7.14 & 3.68 \\  
\hline  
MnBi$_2$Se$_4$ & 4.11 & 12.86 & 68.13 & 22.99 & 22.57 & 60.37 & 0.34 & 0.16 & 0.36 & 0.14 & 7.36 & 10.47 & 4.25 \\ 
\hline    
MnBi$_2$S$_4$ & 3.96 & 12.36 & 80.80 & 27.45 & 26.68 & 71.48 & 0.34 & 0.14 & 0.34 & 0.14 & 8.60 & 11.92 & 4.53 \\ 
\hline\hline
\end{tabular}
}
\label{table:1}    
\end{table*}	

In Figs.~\ref{fig:strain-stress}a and~\ref{fig:strain-stress}b, we show the top and side view of the MnBi$_2$X$_4$ structure, which has a hexagonal structure and consists of seven atomic layers with the sequence of X-Bi-X-Mn-X-Bi-X (X = Te/Se/S) in ABC-type arrangement along the $z$-direction. MnBi$_2$X$_4$ has the $P\overline{3}m1$ space group, and the optimized in-plane lattice parameters $a$ are listed in Table~\ref{table:1}, consistent with previous work~\cite{an2021nanodevices}. Since Te has a larger atomic radius ($\sim 140$ pm) than Se ($\sim 120$ pm) and S ($\sim 100$ pm), the bond length of Bi-Te is naturally longer than that of Bi-Se and Bi-S. Thus, the lattice constant $a=4.36$ \AA\ of MnBi$_2$Te$_4$ is higher than that of MnBi$_2$Se$_4$ (4.11 \AA) and MnBi$_2$S$_4$ (3.96 \AA).

In Figs.~\ref{fig:strain-stress}c-\ref{fig:strain-stress}e, we show the stress-strain curves of MnBi$_2$X$_4$ for $s_{\text{bia}}$, $s_{xx}$, and $s_{yy}$, respectively. The mechanical strength of MnBi$_2$S$_4$ is larger than that of MnBi$_2$Se$_4$ and MnBi$_2$Te$_4$ because S is more electronegative (2.58 on the Pauling scale) than Se (2.55) and Te (2.1). Higher electronegativity leads to stronger bonding interactions~\cite{van2020first}, pulling the atoms closer together (smaller lattice constant). As shown in Fig.~\ref{fig:strain-stress}g, the integral of projected crystal orbital Hamilton population (IpCOHP) of spin-down (dashed lines) and spin-up (solid lines) of MnBi$_2$S$_4$ is more negative values than that of MnBi$_2$Se$_4$ and MnBi$_2$Te$_4$, which indicates stronger bonding due to a larger amount of bonding states being occupied. The spin-up also includes significant anti-bonding contributions near the Fermi energy at 0 eV compared with spin-down. Thus, the bond strength of MnBi$_2$X$_4$ is mainly contributed by the spin-down. 

\begin{figure}[t]
    \centering
    \includegraphics[width=1\linewidth]{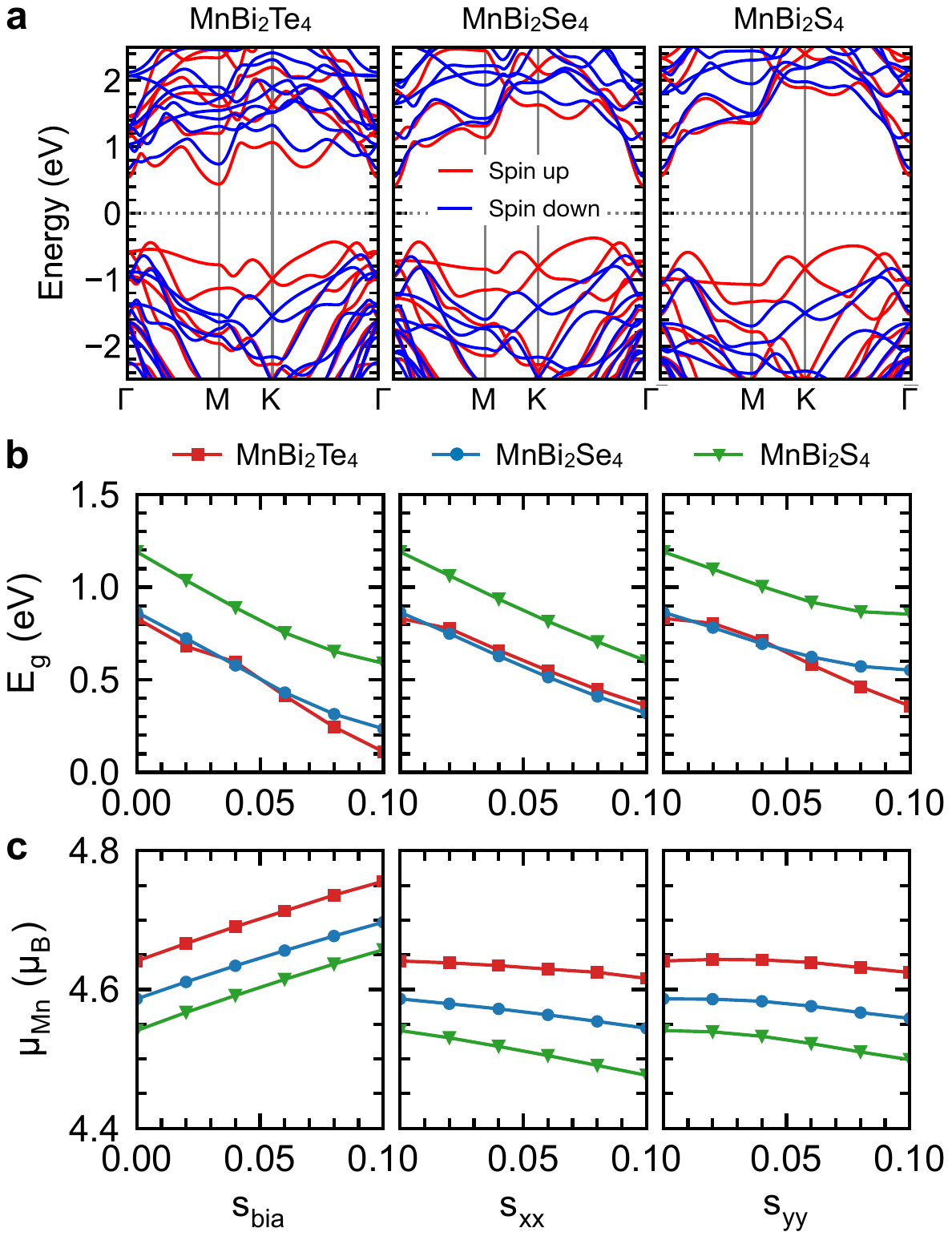}
    \caption{(a) Spin-polarized band structures of MnBi$_2$X$_4$ (X = Te/Se/S) by using DFT+U. (b) Energy band gap $E_g$ and (c) magnetic moment of Mn atom are plotted as a function of biaxial strain ($s_{\text{bia}}$) and uniaxial strains along the $x$- ($s_{xx}$) and $y$-directions ($s_{yy}$).}
    \label{fig:band}
\end{figure}

The maximum values of the stress-strain curves indicate the critical strain $s^c$ and ideal strength $\sigma^i$, which refer to the theoretical limits of a material's mechanical properties before failure~\cite{zhu2010ultra,hung2016intrinsic}. Table~\ref{table:1} lists the values of $s^c$, $\sigma^i$, and other mechanical properties, including the elastic modules $C_{ij}$, Young modulus $Y$, and Poisson ratio $r$. It noted that the parameters in the unit of GPa are scaled by an effective thickness $h$ of 2D MnBi$_2$X$_4$ (see values of $h$ in Table~\ref{table:1}), which is obtained by the optimized structure of the bulk 3D-layered MnBi$_2$X$_4$ by accounting for the van der Waals interaction. $C_{ij}$ are obtained by using the strain-energy calculations (see Fig. S3 in Supporting Information). The ideal strength in the $x$-direction is much larger than that in the $y$-direction ($\sigma^i_{xx}/\sigma^i_{yy}\sim 2.0-2.5$). This is because the X-Bi-X bonds slip with each other in the $y$-direction plane while it is stretched in the $x$-direction plane, as shown in Fig.~\ref{fig:strain-stress}f by the electron localization function (ELF). In Figs. S4a-c, we show the ELF of MnBi$_2$Te$_4$ at the critical strains $s_{\text{bia}}^c=0.16$, $s_{xx}^c=0.27$, and $s_{yy}^c=0.16$, respectively. The bonding structure is also the origin of the ductile fracture for the $s_\text{bia}$ and $s_{yy}$ cases and the brittle fracture for the $s_{xx}$ case. The mechanical strength of MnBi$_2$X$_4$ is comparable to that of flexible 2D materials, such as transition metal dichalcogenides, graphene, and Janus materials~\cite{hung2016intrinsic,van2022effects,van2020first}.

In Fig.~\ref{fig:band}a, we show the projected spin-up and -down electronic bands of MnBi$_2$X$_4$. The conduction band minimum is located at the M point for MnBi$_2$Te$_4$ and $\Gamma$ point for MnBi$_2$Se$_4$ and MnBi$_2$S$_4$, while the valence-band maximum is located between the $\Gamma$ and K points, resulting in the indirect band gap $E_g$ of 0.83, 0.86, and 1.19 eV for MnBi$_2$Te$_4$, MnBi$_2$Se$_4$, and MnBi$_2$S$_4$, respectively. The calculated $E_g=0.83$ eV reproduces the observed $E_g = 0.78\pm 0.1$ eV for MnBi$_2$Te$_4$ by angle-resolved photoemission spectroscopy measurements~\cite{trang2021crossover}. We note that the calculated band structures with spin-orbit coupling (SOC) show $E_g = 0.08, 0.33$, and 0.59 eV for MnBi$_2$Te$_4$, MnBi$_2$Se$_4$, and MnBi$_2$S$_4$, respectively (see Fig. S2 in Supporting Information), which are much smaller than that in the experiment$^{12}$. Since $E_g$ is an essential parameter for the optical properties, in this study, we selected the DFT+U without SOC for the optical calculations. In Fig.~\ref{fig:band}b, $E_g$ of MnBi$_2$X$_4$ is plotted as a function of $s_{\text{bia}}$, $s_{xx}$, and $s_{yy}$, in which $E_g$ decreases almost linearly with increasing strain, and MnBi$_2$X$_4$ remains semiconducting ($E_g > 0$) under strains up to 0.1 (10\%). Since evaluating the optical properties and quantum weight requires the material to remain semiconducting, we set 0.1 as the strain limit for the optical calculations. The optimized lattice parameters of MnBi$_2$X$_4$ at each strain up to 0.1 are given in Table S1 in Supporting Information.

In Fig.~\ref{fig:band}c, we show the magnetic moment of the Mn atom, $\mu_\text{Mn}$, as a function of $s_{\text{bia}}$, $s_{xx}$, and $s_{yy}$, respectively. The total magnetic moment of MnBi$_2$X$_4$ is 5 $\mu_B$ per unit cell, mostly contributed by the Mn atom, which is 4.64, 4.59, and 4.54 $\mu_B$ for MnBi$_2$Te$_4$, MnBi$_2$Se$_4$, and MnBi$_2$S$_4$, respectively, consistent with previous work~\cite{li2019magnetic}. The biaxial strain iso-increases the distance between Mn atoms, resulting in the localized spin state that is preferable for magnetism~\cite{zhuang2016strong}. Thus, the magnetic moment of the Mn atom is increased by increasing biaxial strain. However, for the uniaxial strain, the distance between Mn atoms is decreased in the perpendicular direction by the Poisson ratio. It might lead to the overlap between Mn-$3d$ orbitals along one direction and the low-spin states. Thus, in contrast to the case of $s_{\text{bia}}$, the magnetic moment of the Mn atom is decreased by increasing uniaxial strains $s_{xx}$ and $s_{yy}$.


\begin{figure}[t]
    \centering
    \includegraphics[width=1\linewidth]{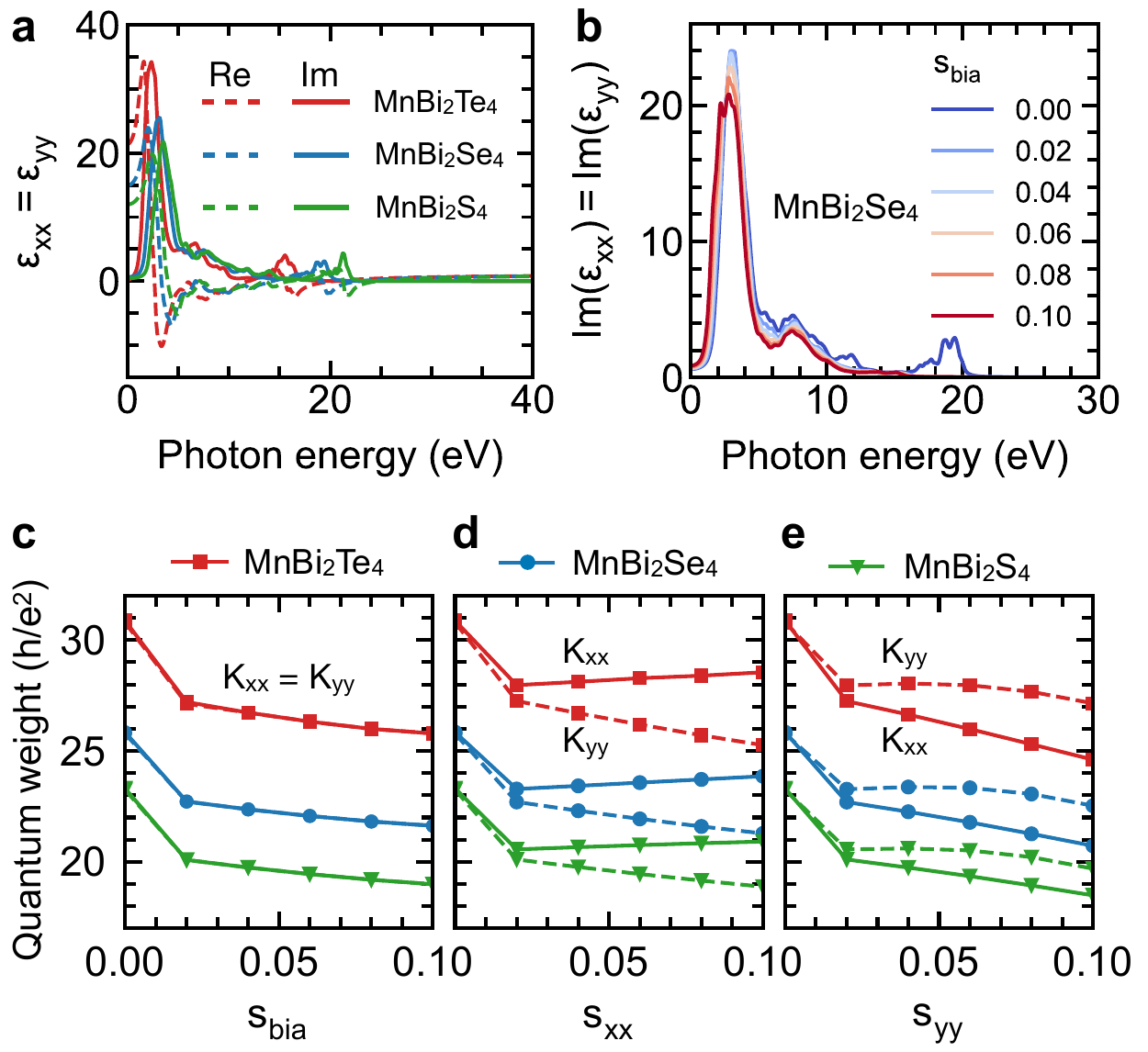}
    \caption{(a) Real $\text{Re}(\varepsilon)$ and imaginary $\text{Im}(\varepsilon)$ parts of in-plane dielectric function of MnBi$_2$X$_4$ (X = Te/Se/S) are plotted as a function of photon energy. (b) $\text{Im}(\varepsilon)$ of MnBi$_2$Se$_4$ is plotted as function of biaxial strain ($s_{\text{bia}}$). Quantum weight $K_{xx}$ and $K_{yy}$ of MnBi$_2$X$_4$ are plotted as a function of (c) $s_{\text{bia}}$ and uniaxial strains along the (d) $x$- ($s_{xx}$) and (e) $y$-directions ($s_{yy}$).}
    \label{fig:optic}
\end{figure}

In order to understand the magnetic configuration, the magnetic anisotropy energy (MAE) of the monolayer MnBi$_2$X$_4$ is calculated by the difference in total energy for the two systems as~\cite{kajale2024field}:
\begin{equation}
    \Delta E = E_\text{SOC}^{m\|z}-E_\text{SOC}^{m\|x},
\end{equation}
where $E_\text{SOC}^{m\|z}$ and $E_\text{SOC}^{m\|x}$ are the total energy including SOC of out-of-plane magnetization (along the $z$ axis) and in-plane magnetization (along the $x$ axis), respectively. In Fig. S5 in Supporting Information, the MAE is plotted as a function of $s_\text{bia}$, $s_{xx}$, and $s_{yy}$. The influence of strain on the MAE is found to be significant for the monolayer MnBi$_2$S$_4$ with biaxial strain, in which MnBi$_2$S$_4$ prefers in-plane magnetization with a negative MAE of $-2.64$ meV at $s_\text{bia}=0.02$ and out-of-plane magnetization with a positive MAE of 2.05 meV at $s_\text{bia}=0.06$.

For a gapped material, such as a semiconductor or an insulator, the quantum metric $g_{\mu \nu}$ is defined by the quantum distance between two states $\psi$ at neighboring momentum $\bm{k}$ and $\bm{k}+\delta\bm{k}$ as $g_{\mu \nu}(\bm{k})\delta k_{\mu}\delta k_{\nu} = 1-|\langle\psi(\bm{k})|\psi(\bm{k}+\delta\bm{k})\rangle|^2$, where $\mu$ and $\nu$ denote the $x$, $y$, or $z$ directions~\cite{provost1980riemannian}. This quantum metric has been directly linked to the real part of the optical conductivity $\text{Re}[\sigma_{\mu\nu}(\omega)]$ as~\cite{onishi2024fundamental,ghosh2024probing,ezawa2024analytic}
\begin{equation}
\int_0^\infty\mathrm{d}\omega\frac{\text{Re}[\sigma_{\mu\nu}(\omega)]}{\omega}=\frac{\pi e^2}{\hbar}\int[\mathrm{d}\bm{k}]\sum_n^{\text{occ}} g_{\mu\nu}^{nn},
\label{eq:1}
\end{equation}
where $\omega$ is the frequency of light, and $n$ is the index of the occupied bands (or valence bands). On the other hand, from the $f$-sum rule, the quantum weight $K_{\mu\nu}$ is defined by~\cite{onishi2024fundamental,hung2024universal}
\begin{equation}
K_{\mu\nu}= \frac{2\hbar}{e^2}\int_0^\infty \frac{\text{Re}[\sigma_{\mu\nu}(\omega)]}{\omega}\mathrm{d}\omega=\frac{2\hbar}{e^2}\int_0^\infty \text{Im}[\varepsilon_{\mu\nu}(\omega)]\mathrm{d}\omega,
\label{eq:2}
\end{equation}
where $\text{Im}[\varepsilon_{\mu\nu}(\omega)]$ is the imaginary part of the dielectric function.
From Eqs. \eqref{eq:1} and \eqref{eq:2}, we have the relationship $K_{\mu\nu}=2\pi\int[\mathrm{d}\bm{k}]\text{Tr}[g_{\mu\nu}]$, where $\text{Tr}[g_{\mu\nu}] = \sum_n g_{\mu\nu}^{nn}$ is the trace of the quantum metric of the occupied band manifold. In other words, $K_{\mu\nu}$ is related to the electron localization length in a gapped material, representing a quantitative parameter of the degree of quantumness of the material~\cite{ghosh2024probing}. Equation~\eqref{eq:2} also tell us that the $K_{\mu\nu}$ can be obtained by integral of $\text{Im}[\varepsilon_{\mu\nu}(\omega)]$ over the entire frequency range~\cite{hung2024universal}.

In Fig.~\ref{fig:optic}a, we show the real $\text{Re}[\varepsilon(\omega)]$ and imaginary $\text{Im}[\varepsilon(\omega)]$ parts of the dielectric function as a function of photon energy $\hbar\omega$ ranging from 0 to 40 eV. Without strain, the in-plane dielectric function of MnBi$_2$X$_4$ is isotropic and $\varepsilon_{xx}=\varepsilon_{yy}$. $\text{Im}[\varepsilon(\omega)]$ reaches its maximum value when the number of available optical transition channels is maximum, and then it steadily decreases to zero at $\hbar\omega$ of 25 eV. Thus, the $\hbar\omega$ up to 40 eV is enough for the cutoff frequency in Eq.~\eqref{eq:2} to calculate the in-plane quantum weights $K_{xx}$ and $K_{yy}$. Since the maximum value of $\text{Im}[\varepsilon(\omega)]$ of MnBi$_2$Te$_4$ is larger than that of MnBi$_2$Se$_4$ and MnBi$_2$S$_4$, $K_{xx}=K_{yy}=30.88$ (in unit of $h/e^2$) of MnBi$_2$Te$_4$ is higher than MnBi$_2$Se$_4$ (25.85) and MnBi$_2$S$_4$ (23.31). These values are comparable with the well-known topology insulator Bi$_2$Te$_3$ (28.87)~\cite{hung2024universal}.

For the biaxial strain $s_\text{bia}$, the maximum $\text{Im}[\varepsilon(\omega)]$ steadily decreases by increasing $s_\text{bia}$ (see Fig.~\ref{fig:optic}b for the MnBi$_2$Se$_4$ case and Fig. S6 for the MnBi$_2$Te$_4$ and MnBi$_2$S$_4$ cases). A redshift (i.e., shifts to lower frequency) is also found on $\text{Im}[\varepsilon(\omega)]$ spectra under $s_\text{bia}$ due to the decreasing energy band gap (see Fig.~\ref{fig:band}b). Thus, the quantum weight decreases linearly with increasing $s_\text{bia}$, except at $s_\text{bia}=0.02$, as shown in Fig.~\ref{fig:optic}c. This behavior occurs because the \( \text{Im}[\varepsilon(\omega)] \) spectra at high frequencies, around 15-20 eV, are suppressed at $s_\text{bia}=0.02$. We obtain $K_{xx}=K_{yy}$ for the $s_\text{bia}$ case because the biaxial strain does not change the in-plane symmetry of MnBi$_2$X$_4$. However, the uniaxial strains $s_{xx}$ and $s_{yy}$
break this symmetry, leading to anisotropic behavior in the quantum weight for uniaxial strain cases (see Figs.~\ref{fig:optic}d and~\ref{fig:optic}e). The anisotropic behavior of $\text{Im}[\varepsilon(\omega)]$ under $s_{xx}$ and $s_{yy}$ are also plotted in Figs. S7-S9 for MnBi$_2$Te$_4$, MnBi$_2$Se$_4$, and MnBi$_2$S$_4$, respectively. 

To summarize, using first-principles calculations, we demonstrate that strain can manipulate the quantumness of the monolayer MnBi$_2$X$_4$ (X = Te, Se, S) through its optical response. The relationship between fundamental physical parameters, such as quantum weight, magnetic moment, and dielectric function, and strain engineering may be useful for designing a quantum device based on topological materials.

\section*{Supporting Information}
See supporting information for spin-polarized band structures of MnBi$_2$Te$_4$ for several Hubbard U-values, band structures of MnBi$_2$X$_4$ with SOC, strain-energy curves of of MnBi$_2$X$_4$, electron localization function of MnBi$_2$Te$_4$ at critical strains, magnetic anisotropy energy, optimized lattice parameters, and imaginary part of in-plane dielectric function of MnBi$_2$X$_4$ (X = Te, Se, S) under strain.

\section*{Acknowledgments}
N.T.H. acknowledges the financial support from the Frontier Research Institute for Interdisciplinary Sciences, Tohoku University. M. L. acknowledges the
Class of 1947 Career Development Chair and the support from R. Wachnik. T. S. acknowledges JSPS KAKENHI (Grant No's. JP23H00159, JP23K17720, JP24H00032) and JST FOREST Program (Grant No. JPMJFR222H).

\section*{References}
%

\end{document}